\documentclass[runningheads]{svmult}

\usepackage{makeidx}   % allows index generation
\usepackage{graphicx}  % standard LaTeX graphics tool
\usepackage{subeqnar}  % subnumbers individual equations
\usepackage{multicol}  % used for the two-column index
\usepackage{physprbb}  % modified textarea for proceedings,
\makeindex             % used for the subject index
\begin{document}
\title*{$\gamma$-ray Burst Remnants: How can we find them?}
\toctitle{$\gamma$-ray Burst Remnants:
\protect\newline How can we find them?}
% allows explicit linebreak for the table of content

\titlerunning{$\gamma$-ray Burst Remnants}
% allows abbreviation of title, if the full title is too long
% to fit in the running head
%
\author{Rosalba Perna}
%
%\Authorrunning{Perna}
% if there are more than two authors,
% please abbreviate author list for running head
%
%
\institute{Harvard-Smithsonian Center for Astrophysics, 60 Garden
street, Cambridge, MA 01238, USA}

\maketitle              % typesets the title of the contribution

\section{Introduction}

By now there is substantial evidence that Gamma-Ray Bursts (GRBs) originate
at cosmological distances  from very powerful
explosions. The interaction between a GRB and 
its surrounding environment has dramatic consequences on the
environment itself. At early times, the strong X-ray UV 
afterglow flux photoionizes
the medium on distance scales on the order of 100 pc~\cite{PL},
destroys dust~\cite{WD}, creates photoionization edges~\cite{MR}. 
These are short-term effects, that occur while the
afterglow is propagating into the medium.

In this contribution, I discuss the {\em long-term} effects resulting
from the interaction between a GRB and its environment. In particular,
in \S 2, I discuss signatures of the emission spectrum produced while
the heated and ionized gas slowly cools and recombines. Besides
photoionizing the medium with its afterglow, a GRB explosion drives a
blast wave which is expected to have a very long lifetime. 
In \S 3, I discuss possible candidates for
such GRB remnants in our own and in nearby galaxies, and  ways to
distinguish them from remnants due to other phenomena, such as
multiple supernova (SN) explosions.

\section{Spectral signatures of a cooling GRB remnant} 

Let us consider a GRB afterglow source which turns on at time $t=0$ and
illuminates a stationary ambient medium of uniform density $n$,
initially neutral and in thermodynamic equilibrium. Perna, Raymond
\& Loeb (\cite{PRL})
computed, as a function of position and time, the
temperature of the plasma, the ionization state of the elements,
and the emission in the most important lines.  The gas is heated up to
$T\sim 10^5$ K, and cools over a time $t \sim 10^5(T/10^5{\rm
K})/(n_e/1~{\rm cm}^{-3})$~yr. This time, 
combined with the GRB rate $\sim
(10^{6-7} f_b~{\rm yr})^{-1}$ per galaxy~\cite{WB}  ($f_b\leq 1$ is the
beaming factor), implies that in every galaxy there is a
non-negligible probability of finding an ionized GRB remnant at any
given time.  In order to identify cooling GRB
remnants and distinguish them from other emitting regions, such as SN
remnants, HII regions, etc., line diagnostics are extremely
useful. Fig. 1 (from~\cite{PRL}) shows some line ratios which are
very sensitive to the type of mechanism by which the gas has been
ionized. For a cooling GRB remnant, they reach
much higher values than in shock heated gas and HII regions.
For example, the ratios [O III] $\lambda 5007/{\rm H}\beta$ 
and [O II] $\lambda 3727/{\rm H}\beta$ are typically smaller than $\sim 5$ 
in shock heated gas, while He II $\lambda 4686/{\rm H}\beta$ is 
on the order of a few tenths.  
Here, they can all be as high as $\sim 100$.
This is a consequence of having a huge mass of gas which is suddenly
photoionized and then let cool.
\begin{figure}[t]
\begin{center}
\includegraphics[width=.5\textwidth]{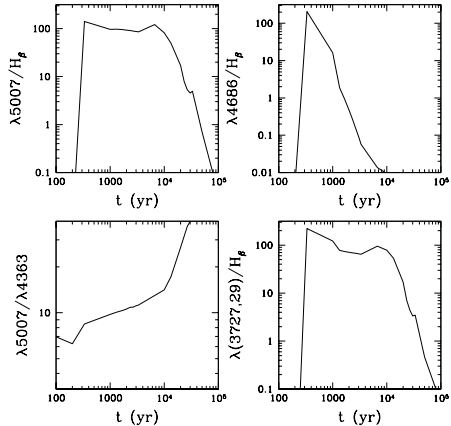}
\end{center}
\caption[]{Some line diagnostics of a cooling GRB remnant of energy $10^{52}$ ergs in
a medium of density $1$ cm$^{-3}$.}
\label{fig1}
\end{figure}
Besides line diagnostics, cooling
GRB remnants can also be identified as they are center-filled with
high ionization lines, and limb-brightened with low-ionization
lines \cite{PRL}. 
The non-relativistic blast wave might be visible separately,
since it does not reach the outer edge of these young photo-ionized
remnants.  Furthermore, the remnants should show evidence for
ionization cones if the prompt or afterglow UV emission from GRBs is
beamed. Therefore, their identification could help constrain the
degree of beaming of the GRB emission.

\section{GRB remnants: hydrodynamic effects}

Besides photoionizing the medium with its afterglow, a GRB explosion
drives a blast wave into the medium. This will 
be washed out by interstellar turbulence 
only after it has slowed down to a velocity of $\sim 10~{\rm
km~s^{-1}}$.  
For a uniform medium of density $n_{1}~{\rm cm^{-3}}$
and an energy of $10^{54}~{\rm ergs}$ deposited in the gas,
this will happen at a radius $R_{\rm kpc}= 0.7~E_{54}^{0.32}
n_{1}^{-0.36}$ after a time $t=2.1~\times~10^7E_{54}^{0.32}
n_{1}^{-0.36}$~yr. 
Given the GRB rate per galaxy~\cite{WB}, 
one estimates that a few such GRB remnants should be
present in every galaxy at any given time. Have we already identified
them?  Maybe. For several decades, 21 cm surveys of spiral galaxies
have revealed the puzzling existence of expanding giant HI supershells~\cite{H}
whose radii  are
much larger than those of ordinary SN remnants and often exceed
$\sim 1$ kpc.  Their estimated ages are in the range $10^6$--$10^8$
years.  Whereas
small shells of radii $\sim 200$--400 pc and energies $\le$ a few $\times
10^{52}$ ergs are often explained as a consequence of the collective
action of stellar winds and supernova explosions originating from OB
star associations~\cite{MK}, the
energy source of the largest supershells is still a subject of debate.
If due to multiple SNe, some of these supershells would require about
$10^4$ SNe to power them; such large OB associations, even though not
impossible, have never been observed in a survey of thousands of HII
regions in nearby galaxies~\cite{KEH}.
\begin{figure}[t]
\begin{center}
\includegraphics[width=.3\textwidth]{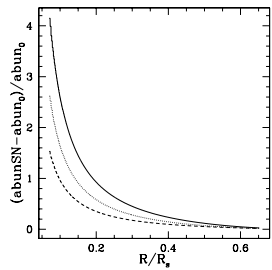}
\end{center}
\caption[]{Metal enhancements for a giant supershell powered by multiple SN explosions. }
\label{eps1}
\end{figure}
The similarity between the expected properties of GRB remnants and
those of HI supershells prompted the suggestion that the energy
source of some of these supershells might actually be GRBs~\cite{EEH,LP}.
This hypothesis could be tested 
based on the fact that SNe inject metals in the ISM in which
they explode. As a result, if a supershell has been powered by
multiple SNe, the abundances of some specific metals in its interior
should be enhanced with respect to the typical values in the ISM
surrounding the shell~\cite{PR}.  Fig. 2 shows the
expected enhancements in the abundances of O, Ne, and
Si for a supershell of energy $E=5\times 10^{53}$ ergs
and age $t=5\times 10^7$ yr that has been powered by multiple SNe.
The enhancements are more pronounced in the inner
regions, as most of
the extra metals are injected at early times due to the shorter
lifetime of the most massive stars.  Such peculiar abundances can be
probed by measuring ratios between X-ray emission lines of two
elements, one of which is more enhanced than the other~\cite{PR}.
Being able to identify which HI supershells have been  produced
by multiple SNe and which ones by GRBs would help
constrain GRB rates and energetics, as well as their location within a
galaxy.

%INDEX%%%%%%%%%%%%%%%%%%%%%%%%%%%%%%%%%%%%%%%%%%%%%%%%%%%%%%%%%%%%%%%
% Please check with the editor of your book whether he plans to
% include a "mutual" subject index - if so, please code your entries
% in the standard syntax. For your own purposes you may print your
% "personal" index by using the following commands:
%
%\clearpage
%\addcontentsline{toc}{section}{Index}
%\flushbottom
%\printindex
%%%%%%%%%%%%%%%%%%%%%%%%%%%%%%%%%%%%%%%%%%%%%%%%%%%%%%%%%%%%%%%%%%%%%

\end{document}